\documentstyle[sprocl]{article}

\def\PRL{\em Phys. Rev. Lett.}

\def\be{\begin{equation}}
\def\ee{\end{equation}}
\def\bea{\begin{eqnarray}}
\def\eea{\end{eqnarray}}
\def\<{{\langle}}
\def\>{{\rangle}}

\def\be{\begin{equation}}
\def\ee{\end{equation}}
\def\ba{\begin{eqnarray}}
\def\ea{\end{eqnarray}}
\def\mref#1{Eq.(\ref{Eq:#1})}

\def\mlab#1{\label{Eq:#1}}

\def\to{\rightarrow}

\def\PR#1#2#3 {{\it Phys. Rev. }{\bf D#1} #2 {(#3)} }
\def\PRL#1#2#3 {{\it Phys. Rev. Lett. }{\bf #1} #2 {(#3)} }
\def\PL#1#2#3 {{\it Phys. Lett. }{\bf #1} #2 {(#3)}  }
\def\AP#1#2#3 {{\it Ann, Phys. }{\bf #1} #2 {(#3)} }
\def\ZP#1#2#3 {{\it Z. Phys. }{\bf #1} #2 {(#3)} }
\def\NP#1#2#3 {{\it Nucl. Phys. }{\bf #1} #2 {(#3)}  }
\def\MPL#1#2#3 {{\it Mod. Phys. Lett.}{\bf #1} #2 {(#3)}  }
\def\NC#1#2#3 {{\it Nuov. Cimm. }{\bf #1} #2 {(#3)}  }
\def\PREP#1#2#3 {{\it Phys. Report }{\bf #1} #2 {(#3)}  }
\def\PROG#1#2#3 {{\it Prog. Theor. Phys. }{\bf #1} #2 {(#3)}   }

\def\sq2{{1\over{\sqrt{2}}}}

\def\g5{\gamma_5}

\newcommand{\matel}[3]{\langle #1|#2|#3\rangle}

\newcommand{\GeV}{\,\mbox{GeV}}

\begin{document}

\begin{flushright}
UND-HEP-99-BIG\hspace*{.2em}07\\
December 1999 \\
%hep-ph/990????\\
\end{flushright}

\vspace{0.5cm}

\title{THE LIFETIMES OF HEAVY FLAVOUR HADRONS - 
A CASE STUDY IN QUARK-HADRON DUALITY 
\footnote{Invited talk given at the 3rd International 
Conference on B Physics and CP Violation, Taipeh, 
Taiwan, Dec. 3 - 7, 1999}}

\author{I. I. BIGI
%\footnote{Invited Lecture given at ...} 
}

\address{Physics Dept.,
Univ. of Notre Dame duLac \\ 
Notre Dame, IN 46556, U.S.A.\\
E-mail: bigi@undhep.hep.nd.edu } 

\maketitle\abstracts{ The status of heavy quark expansions for 
charm and beauty lifetime ratios is reviewed. Taking note of the 
surprising semiquantitative success of this description for 
charm hadrons I interprete the new data on $\tau (D_s)$ and 
re-iterate the call for more precise measurements of  
$\tau (\Xi _c^{0,+})$ and $\tau (\Omega _c)$. A 
slightly larger $B^-$ than $B_d$ lifetime is 
starting to emerge as predicted; the largest lifetime difference in the 
beauty sector, 
namely in $\tau (B_c)$ vs. $\tau (B)$ has correctly been predicted; 
the problem posed by the short $\Lambda _b$ lifetime remains. 
The need for more accurate data also on $\tau (B_s)$ and 
$\tau (\Xi _b^{-,0})$ is emphasized. I discuss quark-hadron 
duality as the central theoretical issue at stake here.} 
%\vskip 5mm
%PACS 11.30.Er, 13.20.Eb, 13.25.Es

%%%%%%%%%%%%
\tableofcontents 
%%%%%%%%%%
%%%%%%%%%%%%%% 
\section{Introduction}
\label{INTRO}
%%%%%%%%%%%%

The lifetime of a hadron represents an 
observable of fundamental as well as practical importance: 
(i) 
Its magnitude reveals whether the decay is driven by 
strong, electromagnetic or weak forces;  
(ii) 
it 
constitutes an essential engineering number for  
translating measured branching ratios 
into widths. 
Yet a strong motivation to {\em measure} a quantity  
does not necessarily imply a need for a precise 
{\em theoretical} description. Furthermore 
we all understand that nothing that is going to happen 
or not happen in the theory of weak lifetimes will make anybody 
abandon QCD as the theory of the strong interactions. After all, 
it is the only game in town after string theory has raised its  
ambition to become the theory of everything rather than merely 
the theory of the strong forces where it had first emerged. 

The 
central theme of my talk will be that developing 
such a theory represents a forum for addressing  
the next frontier in QCD, namely {\em quark-hadron duality} or 
duality for short. 
The concept of duality constitutes an essential element in any 
QCD based 
description and it has been invoked since the early days of the quark 
model. For a long time little progress happened in this area; 
for a violation of duality can be discussed in a meaningful way 
only if one has a {\em reliable} theoretical treatment of 
{\em nonperturbative} effects. 

Let me illustrate that through an 
example. A priori it would be 
quite reasonable to assume that relating the weak decay width 
of a heavy flavour hadron to the fifth power of {\em its} mass rather 
than the heavy {\em quark} mass -- $\Gamma (H_Q) \propto 
M^5(H_Q)$ -- would incorporate boundstate effects as the 
leading nonperturbative corrections (and that is indeed what 
we originally did \cite{MIRAGE}). Only after developing 
a consistent theory for the weak decays of such hadrons through the 
operator product expansion did we realize that such an 
ansatz would violate duality. For it leads to a large 
correction of order $1/m_Q$ -- 
$M^5(H_Q)\simeq (m_Q + \bar \Lambda )^5 \simeq 
m_Q^5(1+5\bar \Lambda /m_Q ) $ -- which is anathema to the OPE 
\cite{HQT}! 

This example already indicates that the study of heavy flavour decays had 
given new impetus to addressing duality: it has provided us 
with new theoretical tools, and it has re-emphasized the need 
to understand the limitations to duality since one aims at 
extracting fundamental KM 
parameters with high numerical accuracy from semileptonic 
decays. 

Nonleptonic transitions provide a rich and multilayered lab to analyze 
duality and its limitations; they can act as a microscope exactly because 
they are thought of containing larger duality violations than semileptonic 
reactions.

%%%%%%%%%%%%%%%%%
\section{Heavy Quark Expansions}
\label{HQE} 
%%%%%%%%%%%%%

The weak decay 
width for a heavy flavour hadron $H_Q$  
into an inclusive final state $f$ can be expressed 
through an 
operator product expansion (OPE) 
\cite{HQT,VICTOR}: 
$$ 
\Gamma (H_Q \to f) = \frac{G_F^2m_Q^5}{192 \pi ^3}
|V_{CKM}|^2  \left[ c_3^{(f)}
\matel{H_Q}{\bar QQ}{H_Q} + 
c_5^{(f)} \frac{\mu _G^2(H_Q)}{m_Q^2}
 + \right. 
$$
\be
\left. + \sum _i c_{6,i}^{(f)} 
\cdot \frac{
\matel{H_Q}{(\bar Q\Gamma _iq)(\bar q\Gamma _iQ)}
{H_Q}}{m_Q^3} + {\cal O}( 1/m_Q^4 ) 
\right] \; , \; \; 
\mlab{MASTER}
\ee 
where 
$\mu _G^2(H_Q) \equiv 
\matel{H_Q}{\bar Q\frac{i}{2}\sigma \cdot GQ}
{H_Q}$. 
\mref{MASTER} exhibits the following important features:
\begin{itemize}
\item  
The expansion involves  
\begin{itemize}
\item 
c-number coefficients $c_i^{(f)}$ calculable within 
short-distance physics; 
\item 
expectation 
values of local operators given by long distance 
physics; their values can be inferred from symmetry 
arguments, other observables, QCD sum rules, lattice studies and quark 
models; 
\item 
inverse powers of the heavy quark mass $m_Q$ 
scaling with 
the known dimensions of the various operators. 
\end{itemize}  
The nonperturbative effects on the decay width -- 
a dynamical quantity -- can thus be expressed through expectation 
values and quark masses. Those being static quantities can be 
calculated with decent reliability. 
\item  
A crucial element of Wilson's prescription for this expansion 
is that it allows a selfconsistent separation of short-distance 
dynamics that is lumped into the coefficients $c_i^{(f)}$ and 
long-distance dynamics that enters through the expectation 
values of local operators. This is achieved through the 
introduction of the {\em auxiliary} scale $\mu$ that enters 
both in the coefficients and the 
matrix elements. As a matter of principle observables have to be 
independant 
of $\mu$ since Nature cannot be 
sensitive to how we arrange our 
computational tasks. In practise, however, $\mu$ has to be chosen 
judiciously for those very tasks: on one hand one would like to 
choose $\mu$ as high as possible to obtain a reliable {\em perturbative}  
expression in powers of $\alpha _S(\mu )$; on the other hand one 
likes to have it as low as possible to evaluate the {\em expectation 
values} in powers of $\mu /m_Q$. This `Scylla and Charybdis' dilemma 
can be tackled by choosing  
$\mu \simeq  1 \; {\GeV}$. 
For simplicity I will not state the dependance on $\mu$ explicitely 
although it is implied.  
\item  
The free quark model 
or spectator 
expression emerges asymptotically 
(for $m_Q \to \infty$) from the $\bar QQ$ operator 
since $\matel{H_Q}{\bar QQ}{H_Q}  
= 1 + {\cal O}(1/m_Q^2)$.
\item   
{\em No} ${\cal O}(1/m_Q)$ contribution can arise in the OPE 
since there is no independant dimension four operator (with colour 
described by a local gauge theory) 
\footnote{The operator $\bar Q i \not D Q$ can be reduced 
to the leading operator $\bar QQ$ through the equation 
of motion.}. This has two important 
consequences: 
\begin{itemize}
\item 
With the {\em leading} nonperturbative corrections 
arising at order $1/m_Q^2$, their size is typically 
around 5 \% in beauty decays. They had not been anticipated 
in the phenomenological descriptions of the 1980's.  
\item 
A $1/m_Q$ 
contribution can arise only due to a massive duality violation. 
Thus one should set a rather high threshold 
before accepting such a statement.  
\end{itemize} 
\item  
Pauli Interference (PI) \cite{GUB}, 
Weak Annihilation (WA) \cite{SONI} 
for mesons and W-scattering (WS) for 
baryons  
arise unambiguously and naturally in order 
$1/m_Q^3$ with WA being helicity 
suppressed \cite{MIRAGE}. They mainly drive the differences in the 
lifetimes of the various hadrons of a given heavy 
flavour.
\end{itemize}

The expectation values of $\bar QQ$ and 
$\bar Q\frac{i}{2}\sigma \cdot GQ$ are known with sufficient 
accuracy for the present purposes from the hyperfine splittings 
and the charm and beauty hadron masses \cite{PRO}. 

The largest uncertainties enter in the expectation values for the 
dimension-six four-fermion operators in order $1/m_Q^3$. For 
{\em mesons} I will invoke approximate {\em factorization at a 
low scale of around 1 GeV}. One should note that factorizable 
contributions at a low scale $\sim$ 1 GeV will be partially 
nonfactorizable at the high scale $m_Q$! 

For {\em baryons} there is no 
concept of factorization, and we have to rely  
on quark model results. 

Below I will discuss mainly hadron-{\em specific} duality violations 
affecting the {\em ratios} between 
different hadrons of a given heavy flavour.

%%%%%%%%%%%%%%%
\section{Lifetimes of Charm Hadrons}
\label{CHARM}
%%%%%%%%%%%%%%

One rough measure for the numerical stability of the 
$1/m_c$ expansion is provided by 
$\sqrt{\mu _G^2(D)/m_c^2} \simeq 0.5$ as an 
effectice expansion parameter which is not small compared 
to one. Obviously one can expect 
-- at best -- a semiquantitative description. 
The mesonic four-quark matrix elements are calibrated by  
$f_D \sim 200$ MeV and $f_{D_s}/f_D \simeq 1.1 - 1.2$. 

On general grounds one expects the following hierarchy 
in lifetimes \cite{BILIC,PRO}:
\be 
\tau (D^+) > \tau (D^0) \sim \tau (D_s^+) \geq 
\tau (\Xi _c^+) > \tau (\Lambda _c^+) > 
\tau (\Xi _c^0) > \tau (\Omega _c) 
\mlab{PATTERNCHARM} 
\ee 
Table \ref{TABLECHARM} shows 
the predictions and data. 
\begin{table}
\begin{tabular} {|l|l|l|l|}
\hline   
 & $1/m_c$ expect. \cite{PRO}& theory comments & data \\ 
\hline 
\hline 
$\frac{\tau (D^+)}{\tau (D^0)}$ & $\sim 2$ & PI dominant & $2.55 \pm 0.034$ 
(updated) \\
\hline 
$\frac{\tau (D_s^+)}{\tau (D^0)}$ & 1.0 - 1.07 & {\em without} WA 
\cite{DS}&
 $1.125 \pm 0.042$  PDG '98\\ 
 & 0.9 - 1.3 & {\em with} WA \cite{DS} & $1.211 \pm 0.017$ \\
 & $1.08 \pm 0.04$ & QCD sum rules 
\cite{HY}& E791, CLEO, FOCUS \\
\hline 
$\frac{\tau (\Lambda _c^+)}{\tau (D^0)}$ & $\sim 0.5$ 
& quark model matrix elements & $0.489 \pm 0.008$ (updated)\\ 
\hline 
$\frac{\tau (\Xi _c^+)}{\tau (\Lambda _c^+)}$ & $\sim 1.3$ & 
ditto & $1.75 \pm 0.36$ PDG '98\\
\hline 
$\frac{\tau (\Xi _c^+)}{\tau (\Xi _c^0)}$ & $\sim 2.8$ & 
ditto & $3.57 \pm 0.91$ PDG '98 \\
\hline 
$\frac{\tau (\Xi _c^+)}{\tau (\Omega _c)}$ & $\sim 4$ & 
ditto & $3.9 \pm 1.7$ PDG '98\\
\hline  
$y = \left. \frac{\Delta \Gamma}{2\Gamma}\right| _{D^0}$ & 
$\leq {\cal O}(1\% )$ & test bed for duality & 
$-6\% \leq y \leq 0.3 \% $ CLEO \\  
\hline 
\end{tabular}
\centering
\caption{Lifetime ratios in the charm sector} 
\label{TABLECHARM} 
\end{table} 
A few comments are in order here: 
\begin{itemize}
\item 
You apply the $1/m_c$ expansion at your own 
risk. It is easy to list reasons why it should fail to reproduce 
even the qualitative pattern expressed in \mref{PATTERNCHARM}. 
However comparing the data with 
the expectations shows agreement even on the semiquantitative level. 
This could be accidental; yet I will explore the possibility 
that it is not. One should note that the longest and shortest lifetimes 
differ by a factor of about twenty! 
\item 
PI is the main engine driving the $D^+ - D^0$ lifetime difference 
as already anticipated in the `old' analysis of 
Guberina et al. \cite{GUB}; the main impact of the HQE 
for this point was to show that WA cannot constitute the leading 
effect and that $BR_{SL}(D^0) \simeq 7 \%$ is consistent with PI being the 
leading effect, see below  
\footnote{$\Gamma (D^+)$ is guaranteed to remain positive if 
the range in momentum over which PI can occur is properly evaluated. 
To put it differently: while one cannot count on obtaining a reliable 
value for $\Gamma (D^+)$, a nonsensical result will arise 
only if one makes a mistake.}. In quoting a lifetime ratio of $\sim 2$ 
I am well aware that the measured value is different from two. Yet that 
numerical difference is within the theoretical noise level: one could 
use $f_D=220$ MeV rather than 200 MeV and WA, which has been 
ignored here, could account for 
10 - 20 \% of the $D^0$ width. 
\item   
Since $\tau (D_s)/\tau (D^0) \simeq 1.07$ can be generated {\em without} WA 
\cite{DS}, 
the `old' data on $\tau (D_s)/\tau (D^0)$ had provided an independant 
test for WA {\em not} being the leading source for 
$\tau (D^0) \neq \tau (D^+)$; it actually allowed for it being quite 
irrelevant. The `new' data reconfirm the first conclusion; at the same time 
they point to WA as a still significant 
process. 
%This provides new impetus to uncover the impact of WA on 
%exclusive channels like semileptonic modes, 3-pion final states etc. 
A recent analysis of WA relying on QCD sum rules 
\cite{HY} is not quite able to 
reproduce the observed lifetime ratio; further analysis along these lines 
is called for. 
\item 
The $1/m_c^2$ contribution controlled by $\mu _G^2(D)$ {\em reduces} the 
semileptonic width common to $D^0$ and $D^+$ mesons; this brings the 
{\em absolute} values observed for $BR_{SL}(D^0)$ and 
$BR_{SL}(D^+)$ in 
line with what is expected when it is mainly PI that generates the 
$D^+ - D^0$ lifetime difference. 
\item 
The description of the {\em baryonic} lifetimes is helped by the 
forgiving experimental errors. More accurate measurements of 
$\tau (\Xi _c^{+,0}, \Omega _c)$ are needed. They might well 
exhibit deficiencies in the theoretical description. 
\item  
{\em Non}universal semileptonic widths -- 
$ 
\Gamma _{SL}(D) \neq \Gamma _{SL}(\Lambda _c) \neq 
\Gamma _{SL}(\Xi _c) \neq \Gamma _{SL}(\Omega _c) 
$ -- 
are predicted with the main effect being {\em constructive} PI in  
$\Xi _c$ and $\Omega _c$ decays; the lifetime ratios among the 
baryons will thus not get reflected in their semileptonic 
branching ratios; one estimates \cite{VOLSL} 
\ba 
BR_{SL}(\Xi _c^0) \sim BR_{SL}(\Lambda _c) 
\;  &\leftrightarrow& \;     
\tau (\Xi _c^0) \sim 0.5 \cdot \tau (\Lambda _c) \\
BR_{SL}(\Xi _c^+) \sim 2.5 \cdot BR_{SL}(\Lambda _c) 
\;  &\leftrightarrow& \;     
\tau (\Xi _c^+) \sim 1.3 \cdot \tau (\Lambda _c) \\ 
BR_{SL}(\Omega _c)  \; &<& 15 \;  \% 
\ea

\item 
On general grounds one expects 
$  
\Delta \Gamma (D^0)/\Gamma (D^0) \leq 
{\rm tg}^2\theta _C \cdot SU(3)_{Fl}$ breaking $\leq 
{\cal O}(0.01 ) 
$.
If the data show that the lifetime difference for the two neutral $D$ 
mass eigenstates is significantly below this bound, one would have 
learned an intriguing lesson on duality.  

\end{itemize}

%%%%%%%%%%%%%%%
\section{Lifetimes of Beauty Hadrons}
\label{BEAUTY}
%%%%%%%%%%%%%%

%%%%%%%%%%%%%%%%%%%
\subsection{Orthodoxy}
%%%%%%%%%%%%%%%%

The numerical stability of the 
$1/m_b$ expansion is characterised by 
$\sqrt{\mu _G^2(B)/m_b^2}$ $ \simeq 0.13 \ll 1$; i.e. 
such an expansion should yield rather reliable numerical 
results. Merely reproducing the qualitative 
pattern would be quite unsatisfactory. I will also use 
$f_B \sim 200$ MeV and $f_{B_s}/f_B \simeq 1.1 - 1.2$. 
Table \ref{TABLEBEAUTY} shows 
predictions 
\cite{BSTONE,PRO} and data. 
\begin{table}
\begin{tabular} {|l|l|l|l|}
\hline   
 & $1/m_b$ expect. & theory comments & data \\ 
\hline 
\hline 
$\frac{\tau (B^+)}{\tau (B_d)}$ & $1+ 0.05 \left( 
\frac{f_B}{200 \; {\rm MeV}}\right) ^2$ & PI in 
$\tau (B^+)$ & $1.066 \pm 0.024$ 
%\cite{LEPAV} 
\\ 
 & & factorization at ~ 1 GeV& \\
\hline 
$\frac{\bar \tau (B_s)}{\tau (B_d)}$ & 
$1.0 \pm {\cal O}(0.01)$   & &
 $0.94 \pm 0.04$ 
%\cite{LEPAV} 
\\ 
 $\frac{\Delta \Gamma (B_s)}{\Gamma (B_s)}$& 
$0.18 \cdot \left( \frac{f_{B_s}}{200 \; {\rm MeV}}\right) ^2$ 
& first calculated in Ref. \cite{DELTAG}  & 
 $\leq$ 0.46  (95 \% C.L.) 
%\cite{LEPAV} 
\\
 & &  & \\
\hline 
$\tau (B_c)$ & $\sim 0.5$ psec  
& largest lifetime difference! & $0.46 \pm 0.17$ psec 
%\cite{BCCDF}
\\ 
\hline 
$\frac{\tau (\Lambda _b)}{\tau (B_d)}$ & 0.9 - 1.0 & 
quark model matrix elements & $0.79 \pm 0.05$ 
%\cite{LEPAV}
\\
\hline 
\end{tabular}
\centering
\caption{Lifetime ratios in the beauty sector} 
\label{TABLEBEAUTY} 
\end{table} 
Again some comments to elucidate these findings:
\begin{itemize}
\item 
The original predictions for the {\em meson} lifetimes, 
which had encountered theoretical criticism 
\cite{NEUBSACH},  
are on the mark. 
(i) A recent lattice study \cite{DIPIERO} finds a result 
quite consistent with the 
original work based on factorization \cite{MIRAGE}:
\be 
\frac{\tau (B^+)}{\tau (B_d)} = 1.03 \pm 0.02 \pm 0.03
\ee  
(ii)  
Sceptics will argue that predicting lifetime ratios close to unity 
is not overly impressive. In response one should point out that the largest 
lifetime difference by far -- $\frac{\tau (B_c)}{\tau (B_d)} \simeq 
\frac{1}{3}$ -- has been 
predicted correctly and that the absence of contributions 
$\sim {\cal O}(1/m_Q)$ had been crucial there! 
\item 
A serious challenge arises from the `short' baryon lifetime. 
In terms of $\Delta \equiv   
1 - \tau (\Lambda _b)/\tau (B_d)$ the data 
can be expressed by 
\be 
\Delta _{\rm experim} = 0.21 \pm 0.05
\mlab{LAMEXP}
\ee
A detailed analysis of 
quark model calculations \cite{BOOST} finds however 
\be 
\Delta _{\rm theor.} = 0.03 - 0.12 
\mlab{LAMTH}
\ee
Reanalyses by other authors agree with \mref{LAMTH} 
\cite{VOLOSHIN} -- as does a {\em pilot} 
lattice study \cite{PILOT}: 
$\Delta _{\rm lattice} = (0.07 - 0.09)$. A recent analysis 
based on QCD sum rules arrives at a significantly larger value:  
$ 
\Delta _{\rm QCD SR} = 0.13 - 0.21
$ \cite{HUANG}.
If true it would remove the problem. However, I would like to 
understand better how 
the sum rules analysis can differ so much from other studies given that it 
still uses the valence quark approximation. 
\item 
An essential question for future studies concerns the lifetimes of the 
beauty hyperons $\Xi _b^{-,0}$. On general grounds one expects 
$\tau (\Xi _b^-) > \tau (\Lambda _b)$, $\tau (\Xi _b^0)$ \cite{PRO}. More 
specifically, using the observed charm hyperon lifetimes and 
$SU(3)$ symmetry a very sizeable effect has been predicted 
\cite{VOLOSHIN}: 
\be 
\frac{\tau (\Xi _b^-) - \tau (\Lambda _b)}{\tau (\Lambda _b)} 
\sim 0.14 - 0.21
\ee

\end{itemize}

%%%%%%%%%%%%%%%%%%%%%%%
\subsection{Heresy}
%%%%%%%%%%%%%%%%%%%%%

As said before, the ansatz $\Gamma (H_Q) \propto 
M(H_Q)^5$ which would yield $\tau (\Lambda _b)/\tau (B_d)$  
$\simeq 0.75$ and therefore has been re-surrected \cite{ALTANO} 
is anathema to the OPE since it would imply  
the nonperturbative corrections to be of order $1/m_Q$! 
The $B^- - B_d$ lifetime difference is still a ${\cal O}(1/m_b^3)$ 
effect. 
\footnote{ 
It has been shown \cite{FIVE} (at least 
for semileptonic transitions) that duality is implemented as follows: 
 `quark phase  space  +  nonperturbative  corrections 
$ 
 \hat = 
$ hadronic phase space +  boundstate effects'!}   

Notwithstanding my employer I am willing to consider heresy, though, 
since it makes some further prediction that differ 
from the OPE findings:  
\be 
\frac{\bar \tau (B_s)}{\tau (B_d)} = 
\left( \frac{M(B_d)}{M(B_s)} \right) ^5 \simeq 0.94 
\ee
\be 
\tau (\Lambda _b) / \tau (\Xi _b^0) / \tau (\Xi _b^-) 
\simeq   1  /  0.85  / 0.85 \; ; 
\mlab{XIHER}
\ee 
the {\em expectation} 
$ M(\Xi _b) - M(\Lambda _b) \simeq M(\Xi _c) - M(\Lambda _c)$ 
has been used in \mref{XIHER}. 
 
On the down side I do not see how such an ansatz can yield 
a correct prediction for $\tau (B_c)$ in a natural way. 

One can also add that such an ansatz helps to understand 
neither the pattern nor the size of the lifetime differences 
in the charm sector. Agreement with the data can be 
enforced, though, by adjusting 
\cite{ALTANO} -- in an ad-hoc fashion I would 
say -- the contributions from PI, WA and WS.

%%%%%%%%%%%%
\section{On Quark-Hadron Duality} 
\label{DUALITY}
%%%%%%%%%%%%%%

%%%%%%%%%%%%%%%
\subsection{General Remarks}
%%%%%%%%%%%%%%

Duality has been an early and somewhat fuzzy element of quark model 
arguments. It can be expressed as follows: "Rates evaluated on the 
{\em parton} level `approximate' observable rates summed over a 
`sufficient number' of {\em hadronic} channels." It was never stated 
clearly 
how good an 
approximation it provided and 
what is meant by `sufficient number'; it was thought, though, that this 
number had to be larger than of order unity. 
%This concept was illustrated for the 
%reaction $e^+ e^- \to \; had$ \cite{PQW}; the first theoretical 
%treatment was then given for the QCD sum rules \cite{SVZ}. 

Heavy quark theory has opened up new theoretical tools as well as 
perspectives onto duality; it demonstrated that duality 
can hold even with one or two channels dominating -- if an 
additional feature like heavy quark symmetry intervenes \cite{SV}. 
This has been demonstrated for semileptonic $b \to c$ decays. 
The goal is to understand better the 
origins of limitations to duality and to develop 
some {\em quantitative} measures for it. The new tools that are 
being brought to bear on this problem are 
(a)  
the OPE; 
(b)  
the so-called small velocity sum rules \cite{OPTICAL} 
and (c)   
the 't Hooft model.

The results obtained so far show 
there are different categories of duality -- 
local vs. global etc. duality -- depending on the amount 
of averaging or `smearing' that is involved 
%\footnote{Global duality should 
%not be identified with what is often referred to as just 
%duality \cite{OPTICAL}.}
 and that duality can neither be universal nor 
exact. 

Duality is typically based on dispersion relations expressing 
observable rates averaged over some kinematical variables through 
an OPE constructed in the Euclidean region. There are 
natural limitations to the accuracy of such an expansion; 
among other things it will have to be truncated. In any case, such 
a power expansion will exhibit {\em no} sensitivity to a term like 
$exp(-m_Q/\Lambda _{QCD})$. Yet upon analytical  
continuation from the 
Euclidean to the Minskowskian domain this  
exponentially suppressed term turns into sin$(m_Q/\Lambda _{QCD})$ -- 
which is not surpressed at all! I.e., the OPE {\em cannot} account 
for such terms that could become quite relevant in Minkowski 
space and duality violations can thus enter through these  
`oscillating' terms; the opening of thresholds provides 
a model for such a scenario.  Actually they will be surpressed somewhat 
like $(1/m_Q^k)\sin (m_Q/\Lambda _{QCD})$ with the power $k$ 
depending on the dynamics in general and the reaction in particular. 
This could produce also a `heretical' $1/m_Q$ contribution 
from a dimension-five operator: 
\be 
\frac{1}{m_Q^2} \sin \left( \frac{m_Q}{\Lambda _{QCD}} \right) = 
{\cal O}(1/m_Q)
\ee 

The colour flow in semileptonic as well as 
nonleptonic spectator decays and in WA is such that duality can 
arise naturally; i.e., nature had to be malicious to create sizeable 
duality violations. Yet in PI -- because it is an interference 
phenomenon -- the situation is much more complex leading to serious 
concerns about the accuracy with which duality can apply here. 

%%%%%%%%%%%%%%%%%
\subsection{'t Hooft Model}
%%%%%%%%%%%%%%%%%

The most relevant features of the 't Hooft model are: 
(1)  
QCD in 1+1 dimensions obviously confines. 
(2)  
It is solvable for $N_C \to \infty$: its spectrum   
of narrow resonances can be calculated as can their wavefunctions. 

Duality can then be probed directly by comparing the width evaluated 
through the OPE with a sum over the `hadronic' resonances appropriately 
smeared over the threshold region: 
\be 
\Gamma _{OPE}(H_Q) \; \; \leftrightarrow \; \; 
\sum _n \Gamma (H_Q \to f_n ) 
\ee 
Such a program has been first pursued using 
{\em numerical} methods; it 
lead to claims that duality violations arise in the total 
width through a $1/m_Q$ term 
\cite{GL1} and quantitatively more massively 
in WA \cite{GL2}. 

However an analytical study has shown that 
neither of these claims is correct: perfect matching of the 
OPE expression with the sum over the hadronic resonances was 
found through high order in $1/m_Q$ \cite{TH1,TH2}. The 
same result was obtained for the more intriguing case of 
PI \cite{TH2,TH3}.

%%%%%%%%%%%%%%%
\section{Summary and Outlook}
\label{SUMMARY}
%%%%%%%%%%%%%%

A mature formalism genuinely based on QCD 
has been developed for describing inclusive nonleptonic 
heavy flavour decays. It can tackle questions that could not 
be addressed before in a meaningful way; even failures can 
teach us valuable lessons on nonperturbative aspects of QCD, 
namely on limitations to duality. 

A fairly successful semiquantitative picture has emerged for 
the lifetimes of {\em charm hadrons} 
considering the wide span characterised by 
$\tau (D^+)/\tau (\Omega _c) \sim 20$; while this might be a 
coincidence, it should be noted: 
\begin{itemize}
\item 
PI provides the leading effect driving the $D^+ - D^0$ lifetime 
difference; this conclusion is fully consistent with the 
absolute value for $BR_{SL}(D^0)$. 
\item 
This year's precise new experimental result 
\be 
\frac{\tau (D_s)}{\tau (D^0)} = 1.211 \pm 0.017
\ee 
confirms this picture: WA is nonleading, though still significant. 
It represents an interesting challemge to find the 
footprint of WA in some classes of exclusive final states. 
\item 
More precise data on the $\Xi _c^{0,+}$ and $\Omega _c$ 
lifetimes are very much needed. Those might  
reveal serious deficiencies in the predictions. It should 
be noted that the semileptonic widths for 
baryons are {\em not} universal!  
\end{itemize}
The scorecard for beauty lifetimes looks as follows: 
\begin{itemize}
\item 
The predictions for $\tau (B^-)/\tau (B_d)$ and even more impressively 
for $\tau (B_c)$ appear to be borne out by experiment. 
\item 
The jury is still out on $\tau (B_s)$. 
\item 
The $\Lambda _b$ lifetime provides a stiff challenge to theory. 
It should be noted that most authors have shown a 
remarkable lack of flexibility in accommodating 
$\tau (\Lambda _b)/\tau (B_d) < 0.9$, which is quite unusual 
in this line of work. Maybe experiment will show more 
flexibility. 
\item 
Accurate data on $\tau (\Xi _b^{-,0})$ will be essential to 
celebrate success or diagnose failure.

\end{itemize}
Having developed a theoretical framework for treating nonperturbative 
effects, we can address the issue of duality violations. While we 
have begun to understand better their origins, we have not (yet) found 
a model {\em theory} that could explain the short $\Lambda _b$ lifetime as  
a duality limitation.    

There are, of course, different layers of failure conceivable and the 
lessons one would have to draw: 
\begin{itemize}
\item 
A refusal by the data to move up the value of $\tau (\Lambda _b)$ 
{\em could} be interpreted as showing that the quark model provides a 
very inadequate tool to estimate baryonic expectation values. 
\item 
A low value of the average $B_s$ lifetime -- say 
$\bar \tau (B_s) < 0.96 \tau (B_d)$ -- had to be seen as a very 
significant limitation to duality.

\end{itemize}
Clearly there is a lot we will learn from future data on 
lifetimes and other inclusive rates -- one way or the 
other.  

%%%%%%%%%%%

\vskip 3mm  
{\bf Acknowledgements} 
\vskip 3mm  
This truly inspiring and enjoyable meeting organized by 
Profs. H.-Y. Cheng and G. Hou clearly wetted the appetite of participants 
for the next incarnation of this series, namely BCP 4. 
I am grateful to my collaborators Profs. M. Shifman, 
N. Uraltsev and A. Vainshtein for generously sharing 
their insights with me. 
This work has been supported by the NSF under the grant 
PHY 96-0508.

%%%%%%%%%%%%%%%%%%%%%%%%%%%%%%%


\begin{thebibliography}{99}

\bibitem{MIRAGE}
I. Bigi, N. Uraltsev, {\em Phys. Lett.} {\bf B 280} 
(1992) 271. 

\bibitem{HQT} 
For a review with references to the earlier 
literature, see: I.I. Bigi, M. Shifman, N.G. Uraltsev, 
{\em Annu. Rev. Nucl. Part. Sci.} {\bf 47} (1997) 591. 

\bibitem{VICTOR} 
For a somewhat different approach, see: 
V. Chernyak, {\em Nucl. Phys.} 
{\bf B 457} (1995) 96.

\bibitem{GUB} 
B. Guberina, S. Nussinov, R. Peccei, R. R\" uckl, {\em Phys. Lett.} {\bf B 89} 
(1979) 261. 

\bibitem{SONI} 
M. Bander, D. Silvermann, A. Soni, {\em Phys. Rev. Lett.} {\bf 44} (1980) 7; 
H. Fritzsch, P. Minkowski, {\em Phys. Lett.} {\bf B 90} 
(1980) 455; W. Bernreuther, O. Nachtmann, B. Stech, {\em Z. Phys.} {\bf C 4} (1980) 257; 
I. Bigi, {\em Z. Phys.} {\bf C 5} (1980) 313. 

\bibitem{BILIC} 
N. Bilic, B. Guberina et al., J. Trampetic, 
{\em Nucl. Phys.} {\bf B 248} 
(1984) 33; M. Shifman, M. Voloshin, {\em Sov. J. Nucl. Phys.} 
{\bf 41} (1985) 120.  

\bibitem{PRO}
For a review with references to the earlier 
literature, see: B. Bellini, I. Bigi and P. Dornan, {\em Phys. Rep.}  
{\bf 289} (1997) 1. 



\bibitem{DS} 
I.I. Bigi, N.G. Uraltsev, {\em Z. Phys.} {\bf C 62} (1994) 623.

\bibitem{HY}
H.-Y. Cheng, K.-C. Yang, {\em Phys. Rev.} {\bf D 61} 
(2000) 014008. 

\bibitem{VOLSL} 
M. Voloshin, {\em Phys. Lett.} {\bf B 385} 
(1996) 369.    

\bibitem{DELTAG} 
M. Voloshin et al., 
{\em Sov.J.Nucl.Phys.} {\bf 46} (1987) 112

\bibitem{BSTONE} 
I.I. Bigi, B. Blok, M. Shifman, N. Uraltsev,  
A. Vainshtein, in: `$B$ Decays', S. Stone (ed.), 
World Scientific, 1994, Revised 2nd Ed., p. 132.  

\bibitem{NEUBSACH}
M.Neubert, C. Sachrajda, 
{\em Nucl. Phys.} 
{\bf B 483} (1997) 339.  

\bibitem{DIPIERO} 
M. Di Pierro, C. Sachrajda, 
{\em Nucl. Phys.} 
{\bf B 534} (1998) 373. 

\bibitem{BOOST}
N.G. Uraltsev, 
{\em Phys. Lett.} {\bf B 376} (1996) 303. 

\bibitem{VOLOSHIN}
M. Voloshin, preprint hep-ph/9901445; 
B. Guberina, B. Melic, H. Stefancic, preprint hep-ph/9907468; 
B. Melic, 
these Proceed.

\bibitem{PILOT} 
M. Di Pierro, C. Sachrajda, C. Michael, preprint 
hep-lat/9906031.

\bibitem{HUANG} 
C.-S. Huang, C. Liu, S.-L. Zhu, preprint hep-ph/9906300.

\bibitem{ALTANO} 
G. Altarelli, G. Martinelli, S. Petrarca, F. Rapuano, 
{\em Phys. Lett.} {\bf B 382} (1996) 409. 


\bibitem{FIVE} 
I. Bigi, M. Shifman, N. Uraltsev, A. Vainshtein, 
{\em Phys. Rev.} {\bf D56} (1997) 4017. 

\bibitem{SV} 
M. Voloshin, M. Shifman, {\em Sov. J. Nucl. Phys.} {\bf 47} 
(1988) 511. 

\bibitem{OPTICAL} 
I. Bigi, M. Shifman, N. Uraltsev, A. Vainshtein, 
{\em Phys. Rev.} {\bf D52} (1995) 196. 

\bibitem{GL1}
B. Grinstein, R. Lebed, {\em Phys. Rev.} {\bf D 57} 
(1998) 1366. 

\bibitem{GL2}
B. Grinstein, R. Lebed, {\em Phys. Rev.} {\bf D 59} 
(1999) 054022. 



\bibitem{TH1} 
I. Bigi, M. Shifman, N. Uraltsev, A. Vainshtein, 
{\em Phys. Rev.} {\bf D59} (1999) 054011. 





\bibitem{TH2} 
I.I. Bigi, N.G. Uraltsev, 
{\em Phys. Rev.} {\bf D60} (1999) 114034. 

\bibitem{TH3} 
I.I. Bigi, N.G. Uraltsev, {\em Phys.Lett.} {\bf B 457} (1999) 163.

%\bibitem{URALTSEV} 
%M. Voloshin, M. Shifman, N. Uraltsev, V. Khoze, 
%{\em Sov. J. Nucl. Phys.} {\bf 46} (1987) 112. 


















\end{thebibliography}
\end{document}